\documentclass[12pt]{iopart}
\usepackage{graphicx}
\usepackage{color}
\usepackage{subfigure}

\begin{document}

\title{Anomalous phase shift in a twisted quantum loop}

\author{Hisao Taira$^{\rm 1,3)}$ and Hiroyuki Shima$^{\rm 2,3)}$}

\address{${}^{\rm 1)}$ Division of Applied Physics, Graduate School of Engineering, Hokkaido University, Sapporo, Hokkaido, 060-8628, Japan }

\address{${}^{\rm 2)}$ Division of Applied Physics, Faculty of Engineering, Hokkaido University, Sapporo, Hokkaido, 060-8628, Japan }

\address{${}^{\rm 3)}$ Department of Applied Mathematics 3, 
LaC${\rm \grave{a}}$N, Universitat Polit${\rm \grave{e}}$cnica de Catalunya 
(UPC), Barcelona 08034, Spain}

\ead{taira@eng.hokudai.ac.jp}
\begin{abstract}
Coherent motion of electrons in a twisted quantum 
ring is considered to explore the effect of torsion inherent to the 
ring. 
Internal torsion of the ring composed of helical atomic 
configuration yields a non-trivial quantum phase shift in 
the electrons' eigenstates. This torsion-induced phase shift 
causes novel kinds of persistent current flow and an Aharonov-Bohm 
like conductance oscillation. 
The two phenomena can occur even when no magnetic 
flux penetrates inside the twisted ring, thus being in complete 
contrast with the counterparts observed in untwisted rings. 
\end{abstract}

\maketitle

\section{Introduction}
There exist two classes of geometric effects relevant to coherent 
motion of electrons in low-dimensional systems. 
The first is the occurrence of effective potential fields originating 
from geometric shape of the system 
\cite{1,2}. 
Suppose that electrons are confined in a low-dimensional curved 
geometry such as a quasi-one-dimensional (1D) twisted wire or a 
two-dimensional curved surface. 
Provided the curved shape is sufficiently thin, quantum excitation 
energies in the transverse direction are much larger than those in 
the tangential direction so that the electron's motion is described 
by an effective Hamiltonian of reduced dimension. 
It has been shown that the effective Hamiltonian involves effective 
scalar \cite{1,2,3} and/or vector 
\cite{torsion1,torsion2,torsion3} potentials whose magnitudes 
depend on the local geometry of the system. 
The presence of the effective fields implies that, for example, electrons moving along a thin twisted 
wire feel an effective magnetic field; thus, the electrons will 
exhibit a quantum 
phase shift whose magnitude is determined by the internal torsion 
of the wire. 
Physical consequences of such the geometry-induced potential fields 
have been explored with a great interest in the last 
decade \cite{4,5,6,9,11,13,15,16,17,19,20,21,22,23,pra,pra2},
owing to the technological progress that enables to 
fabricate exotic nanostructures having curved geometry 
\cite{24,25,26,27,28,29,30,31,32}. 

The second geometric effect arises in the cyclic motion of electrons
along a closed loop. 
Generally when an electron travels along a closed loop, it acquires 
a memory of the cyclic motion as a geometric phase in the 
wavefunction \cite{phase}. 
The most important manifestation of the geometric phase is the 
Aharonov-Bohm (AB) effect \cite{33,34,35}; it occurs when an 
electron goes round in 
a loop encircling magnetic flux. 
It is known that the AB effect has many analogues both in quantum 
physics and beyond, which evidences the relevance of the effect to 
diverse fields in physics 
\cite{37,38,39,41,43,47,48,49,50,51,52}. 

The existence of the above two geometric effects implies 
intriguing phenomena peculiar to quasi-1D twisted systems. 
The phase shift induced by torsion is, for instance, 
expected to cause a novel class of persistent current that flows 
along a closed loop of a twisted wire. 
This persistent current flow is novel in the sense that no magnetic 
flux needs to penetrate inside the loop; this feature is distinctive 
compared with usual persistent current in a non-twist quantum loop 
\cite{53,54,55,56,57,58,59,60}. 
In addition, the torsion-induced phase shift should be manifested 
in the conductance of the twisted ring. 
Constructive and destructive interference between the electrons will 
give rise to a conductance modulation as similar to the ordinary AB 
oscillation \cite{33,34,35} ; nevertheless, in the twisted ring, 
the modulation 
(if it occurs) is driven by the torsion instead of penetrating 
magnetic flux. 
To examine the possibility of these two phenomena is interesting 
from viewpoints of both mathematical physics and nanoscale sciences. 

In this article, we present a thorough mathematical derivation of 
the torsion effect on the coherent 
electron transport through a twisted quantum ring. 
We suppose a closed loop of a quantum wire having helical 
atomic configuration, and consider how the 
internal torsion of the configuration results in a non-trivial phase 
shift in the electron's wavefunction. 
Two physical consequences of 
the phase shift are reviewed and reconsidered: they are called the torsion-induced 
persistent current \cite{taira} and the flux-free AB effect \cite{taira2}. 
Both phenomena take place in the absence of magnetic flux 
penetrating 
inside the ring, which is in complete contrast with the ordinary 
counterparts observed in untwisted quantum rings. 
Further detailed analysis is performed to make clear similarities and differences between the phase shift behaviors of twisted and untwisted rings.
\section{Schr\"odinger equation of a quasi-1D twisted system}
The Hamiltonian of a particle moving in a twisted system can be 
formulated via three different approaches. 
The first approach established by Mitchell \cite{torsion2} 
leads us to the Hamiltonian 
that is applicable to arbitrary dimensional twisted systems. 
The second one, suggested by Magarill and ${\rm \acute{E}}$ntin 
\cite{torsion3}, assumes a quasi-1D system having a finite thickness. 
The thickness allows excitations of mobile electrons in the transverse 
direction. 
The last approach, which also assumes a quasi-1D system, is based on 
a simplification that electrons reside in the lowest energy eigenstate 
of the transversal motion \cite{torsion1}. 
The last one applies to sufficiently thin quantum systems with internal 
torsion and is relatively simple rather than other two methods. 
In this section, we give a brief review of the last formulation 
based on which we will deduce the torsion-induced phase shift in a 
sufficiently thin twisted ring.

\begin{figure}[t]
\begin{center}
\includegraphics[width=8cm]{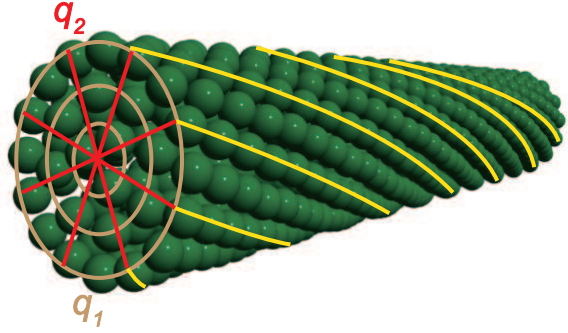}
\caption{\label{fig:fig1} 
Helical atomic configuration consisting of a twisted quantum wire with 
a circular cross section. 
Internal torsion of the wire is defined by the rotation rate of the 
cross section along the wire axis. 
}
\end{center}
\end{figure}

Figure \ref{fig:fig1} illustrates the helical atomic configuration 
that consists of a twisted quantum wire. 
Internal torsion of the wire is defined by the rotation rate of the 
cross section along the central axis $C$. 
The axis $C$ of the wire is parametrized by $q_0$. 
In addition, we introduce curvilinear coordinates $(q_0, q_1, q_2)$ 
such that the $q_1$-$q_2$ plane perpendicular to $C$ rotates along the 
axis with the same rotation rate as that of helical atomic 
configuration (see figure \ref{fig:fig1}). 
A point on $C$ is represented by the position vector 
$\bi{r}=\bi{r}(q_0)$. 
Similarly, a point close to $C$ is given by
\begin{eqnarray}
\bi{R} = \bi{r}(q_0) + q_1\bi{e}_1(q_0) +q_2\bi{e}_2(q_0), 
\label{eq:position}
\end{eqnarray}
where $\bi{e}_1, \bi{e}_2$ are orthogonal unit 
vectors span the cross section. 
$\bi{e}_0$ is defined by 
$\bi{e}_0 \equiv \partial_0 \bi{R}(q_0)$ with the notation 
$\partial_i \equiv \partial/\partial q_i$. 
Since $\bi{e}_0$ is tangential to $C$, the set 
$(\bi{e}_0, \bi{e}_1, \bi{e}_{2})$ composes an orthogonal 
triad. 
Using the helical coordinate system, the Hamiltonian of the system is 
written by \cite{61}
\begin{eqnarray}
H=-\frac{\hbar^2}{2m^*}\sum_{i,j=0}^{2}\frac{1}{\sqrt{g}}\partial_i
\left(\sqrt{g}g^{ij}\partial_j \right)+V.
\label{eq:syure1}
\end{eqnarray}
Here, $m^*$ is the effective mass of electrons and 
\begin{eqnarray}
g={\rm det} [g_{ij}], \ \ g_{ij} = \partial_i\bi{R}\cdot\partial_j \bi{R},\ \ g^{ij}=g_{ij}^{-1}.
\label{eq:metric}
\end{eqnarray}
The term $V$ in equation (\ref{eq:syure1}) represents a strong 
confining potential that constraints the transverse motion of 
electrons' motion within the circular cross section of the radius $d$. 

Let us introduce an important geometric parameter, the internal
torsion $\tau$, defined by 
\begin{eqnarray}
\tau=\bi{e}_2 \cdot \partial_0 \bi{e}_1. 
\label{eq:torsion}
\end{eqnarray}
It quantifies the rotation rate of $\bi{e}_1$ (and $\bi{e}_2$) that 
rotates in a helical manner along $C$. 
In terms of $\tau$ and $\kappa_a = \mbox{\boldmath $e$}_0 \cdot \partial_0 \mbox{\boldmath $e$}_a$, $g_{ij}$ is rewritten by
\begin{eqnarray}
       \begin{array}{l}
g_{00}=\gamma^4+\tau^2\left(q_1^2+q_2^2\right), \\
\vspace{1.0mm}
g_{01}=g_{10}=-\tau q_2, \\
\vspace{1.0mm}
g_{02}=g_{20}=\tau q_1,  \\
\vspace{1.0mm}
g_{ab}=\delta_{ab}, \ \ [a,b=1,2],
       \end{array}
\label{eq:metric2} 
\end{eqnarray}
where $\gamma=(1-\kappa_aq_a)^{1/2}$ (the Einstein convention for repeated indices was used.). 
The elements $g^{ij}$ are those of the $3\times3$ matrix $[g^{ij}]$ 
inverse to $[g_{ij}]$; we can prove that \cite{61}
\begin{eqnarray}
 \begin{array}{l}
g^{00}=\gamma^{-4}, \\  
g^{01}=g^{10}=\gamma^{-4}\tau q_2, \\  
g^{02}=g^{20}= -\gamma^{-4}\tau q_1, \\
g^{ab}=\delta_{ab}+\gamma^{-4}\tau^2\left[ 
\left( q_1^2+q_2^2\right)\delta_{ab}-q_a q_b \right].
 \end{array}
\label{eq:metric3}
\end{eqnarray}
Substituting the results into equation (\ref{eq:syure1}), 
we obtain 
\begin{eqnarray}
H=-\frac{\hbar^2}{2m^*}\left\{\frac{1}{\gamma}\partial_0\frac{1}{\gamma^2}\partial_0\frac{1}{\gamma}+\partial_1^2+\partial_2^2-\frac{1}{\gamma}(\partial_1^2+\partial_2^2)\gamma+ \right.\nonumber \\
\left.\ \ \ \ \ \  \frac{1}{\gamma}\left[\partial_0\tau q_b \frac{1}{\gamma^2}\partial_a + \partial_a\frac{1}{\gamma^2}q_b\tau \partial_0-\left(\partial_0\tau q_a \frac{1}{\gamma^2}\partial_b + \partial_b\frac{1}{\gamma^2}q_a\tau \partial_0\right)\right]\frac{1}{\gamma} + \right. \nonumber \\
\left.\ \ \ \ \ \  \tau^2\frac{1}{\gamma}\partial_a\left[\left(q_1^2+q_2^2\right)\delta_{ab}-q_aq_b\right]\frac{1}{\gamma^2}\partial_b\frac{1}{\gamma}\right\}+V.
\label{eq:generalsyure}
\end{eqnarray}
%
In equation (\ref{eq:generalsyure}), we used the summation convention 
with respect to repeated subscripts: $a,b=1,2$.

For the sake of analytic arguments, we assume that torsion and curvature
of the wire is 
sufficiently smooth and small so that 
$(\kappa_1^2+\kappa_2^2)^{1/2}d \ll1$ and $\tau d \ll 1$. 
Under these conditions, the Hamiltonian (\ref{eq:generalsyure}) is reduced to 
\begin{eqnarray}
H= -\frac{\hbar^2}{2m^*} \left[\left( \partial_1^2 + \partial_2^2 \right)+ \left(\partial_0 - \frac{i \tau L}{\hbar} \right)^2
+\frac{\kappa^2}{4} \right] +V. 
\label{eq:syure2}
\end{eqnarray}
The operator $L\equiv -i\hbar(q_1 \partial_2 - q_2 \partial_1)$
measures the angular momentum of electrons moving in the cross section.
Eigenfunctions of the Hamiltonian (\ref{eq:syure2}) are assumed 
to have the form
\begin{eqnarray}
\phi(q_0, q_1, q_2)=\psi(q_0) \sum^N_{j=1} c_j u_j(q_1, q_2). 
\label{eq:separationvariables}
\end{eqnarray}
Here, $u_j(q_1,q_2)$ are $N$-fold eigenfunctions in the cross section 
and $\psi(q_0)$ describes the axial motion of electrons along the 
twisted wire.
From equations (\ref{eq:syure2}) and (\ref{eq:separationvariables}), 
we can prove that $\psi(q_0)$ obeys the effective one-dimensional 
Schr\"odinger equation such as 
\begin{eqnarray}
-\frac{\hbar^2}{2m^*} \left[\left( \partial_0- \frac{i\tau \langle L \rangle}{\hbar} \right)^2
+\frac{\kappa^2}{4} - \frac{\tau^2}{\hbar^2} \left( \langle L^2 \rangle - \langle L \rangle^2 \right)\right]\psi(q_0) = \epsilon \psi(q_0).
\label{eq:1dsyure2}
\end{eqnarray}
The angular brackets $\langle \cdots \rangle$ indicate to take an 
expectation value with respect to the cross-sectional eigenfunctions 
$u_j(q_1,q_2)$ that are degenerate in general. 
The product $\tau \langle L \rangle/\hbar$ appearing in equation 
(\ref{eq:1dsyure2}) gives rise to a quantum phase 
shift in the wavefunction $\psi(q_0)$, as discussed in detail 
in the next section. 
\section{Torsion-induced phase shift}
\begin{figure}[t]
\begin{center}
\includegraphics[width=8cm]{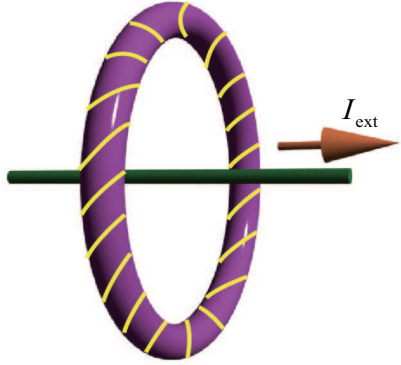}
\caption{\label{fig:fig2} 
Twisted quantum ring encircling the external 
current flow $I_{\rm ext}$. 
}
\end{center}
\end{figure}
This section is devoted to formulate the 
torsion-induced phase shift in a twisted quantum ring. 
The ring radius $R$ ($\gg d$) 
is assumed to be constants throughout the ring. 
Under this assumption, the eigenfunction of the differential 
equation (\ref{eq:1dsyure2}) is given by
\begin{eqnarray}
\psi(q_0) = c\exp\left(-ikq_0 -i \frac{\tau}{\hbar} \int_0^{q_0}
\langle L \rangle dq_0'\right) 
\label{eq:solution}
\end{eqnarray}
with a normalization constant $c$. 
The second term in the parenthesis in equation (\ref{eq:solution}), 
$-(i\tau/\hbar)\int_0^{q_0} \langle L \rangle dq_0'$, 
plays the role of the torsion-induced phase shift. 
The degree of the shift depends on the 
internal torsion $\tau$ of the system, as the name implies.

It follows from equation (\ref{eq:solution}) that a nonzero value of 
$\langle L \rangle$ is necessary for the phase shift 
to occur. 
One possible setup to obtain a finite $\langle L \rangle$ is 
illustrated in figure \ref{fig:fig2}. 
The ring is threaded by an external current flow $I_{\rm ext}$ that 
generates a magnetic field $B$ in the direction tangential to the ring. 
The magnitude of the magnetic field reads $B=\mu_0 I_{\rm ext}/\ell$ 
with $\ell=2\pi R$, where $\mu_0$ is the permeability of vacuum. 
Due to the presence of $B$, the cross-sectional angular momentum 
operator $L$ takes the form of
\begin{eqnarray}
L_B=-i \hbar\frac{\partial}{\partial \theta} -\frac{eBr^2}{2}, 
\label{eq:angularoperator}
\end{eqnarray}
where we used the polar coordinates ($r,\theta$) with respect to 
the circular cross section. 
We shall see below that the expectation value $\langle L_B \rangle$ 
with respect to the cross sectional eigenstates has a nonzero value 
when the current flow $I_{\rm ext}$ is injected. 

The electrons' motion in the cross section subjected to $B$ is 
described by
\begin{eqnarray}
\frac{1}{2m^*}\left\{\frac{1}{r}\left(-i\hbar\frac{\partial}{\partial r}\right)\left[r\left(-i\hbar\frac{\partial}{\partial r}\right)\right]+\left[\frac{1}{r}\left(-i\hbar\frac{\partial}{\partial \theta}\right)+\frac{eBr}{2}\right]^2\right\}u(r,\theta) \nonumber \\
+ V(r)u(r,\theta) = E_{\perp} u(r,\theta) 
\label{eq:syurecross-1} 
\end{eqnarray}
with the transversal energy $E_{\perp}$. 
It is reasonable to set the confining potential $V(r)$  be a parabolic 
well centered at $r=0$, $V(r)=m^*\omega_p^2r^2/2$ \cite{dot,dot2}, 
where $\omega_p$ characterizes the steepness of the potential. 
Then equation (\ref{eq:syurecross-1}) is rewritten as 
\begin{eqnarray}
-\frac{\hbar^2}{2m^*}\left[\frac{1}{r}\frac{\partial}{\partial r}\left(r\frac{\partial}{\partial r}\right) + \frac{1}{r^2} \frac{\partial^2}{\partial \theta^2}\right]u(r,\theta) \nonumber \\
+ \left(-\frac{i\hbar\omega_c}{2}\frac{1}{r}\frac{\partial}{\partial \theta} + \frac{m^*}{8}\omega_c^2r^2+\frac{m^*}{2}\omega_p^2r^2\right) u(r,\theta)= E_{\perp} u(r,\theta), 
\label{eq:syurecross} 
\end{eqnarray}
where $\omega_c=eB/m^*$ is the cycrotron frequency. 
Substituting 
$u(r,\theta)=e^{im\theta}t_{m}(r)/\sqrt{2\pi},\ (m=0,\pm 1,\pm2,\ldots)$ into equation (\ref{eq:syurecross}), we obtain
\begin{eqnarray}
-\frac{\hbar^2}{2m^*}\left[\frac{1}{r}\frac{\partial}{\partial r}\left(r\frac{\partial}{\partial r}\right) - \frac{m^2}{r^2}\right]t_{m}(r) \nonumber \\
+ \left[\frac{m\hbar\omega_c}{2}+\frac{m^*}{2}\left(\omega_p^2+\frac{\omega_c^2}{4}\right)r^2\right] t_{m}(r)= E_{m} t_{m}(r). 
\label{eq:syurecross2} 
\end{eqnarray}
To solve equation (\ref{eq:syurecross2}), we introduce a new variable 
defined by 
\begin{eqnarray}
\rho =\frac{r^2m^*\Omega}{\hbar}, \ \ {\rm where} \ \ \Omega=\sqrt{\omega_p^2+\left(\frac{\omega_c}{2}\right)^2}. 
\label{eq:rho}
\end{eqnarray}
Using the relation 
\begin{eqnarray}
\frac{\partial}{\partial r}=\frac{2rm^* \Omega}{\hbar} \cdot \frac{ \partial}{\partial\rho},  
\label{eq:bibun}
\end{eqnarray}
we can simplify equation (\ref{eq:syurecross2}) as 
\begin{eqnarray}
\rho \frac{d^2 t_m}{d \rho^2}+\frac{dt_m}{d\rho}-\left[\frac{\rho}{4}+\frac{1}{4\Omega}\left(m\omega_c-\frac{2E_m}{\hbar}\right)+\frac{m^2}{4\rho}\right]t_m=0. 
\label{eq:syurecross3} 
\end{eqnarray}
Its solution is given by \cite{Fock,Darwin} 
\begin{eqnarray}
t_m(r)=\displaystyle\sqrt{\frac{2m^*\Omega n!}{\hbar(|m|+n)!}}\exp\left(-\frac{m^*\Omega}{2\hbar}r^2\right)\left(\sqrt{\frac{m^*\Omega}{\hbar}}r\right)^{|m|} \nonumber \\
\hspace{2.0cm} \times L_n^{|m|}\left(\frac{m^*\Omega}{\hbar}r^2\right), \label{eq:syurecross4} 
\\
E_m\ \ \ =\displaystyle(2n+|m|+1)\hbar \Omega + \frac{m}{2}\hbar \omega_c.\ \ \ \ (n=0,1,2,\ldots) \nonumber
\end{eqnarray}
Here, $L_n^{|m|}$ is the associated Laguerre polynomial defined by 
\cite{61}
\begin{eqnarray}
L_n^{|m|}(x)=\frac{d^m}{dx^m}\sum_{a=0}^{n}(-1)^a\frac{(n!)^2}{(a!)^2(n-a)!}x^a.
\end{eqnarray}

We are ready to evaluate the explicit form of $\langle L_B \rangle$. 
From equation (\ref{eq:syurecross4}), we see that all eigenstates 
in the cross section are labeled by the integers $n$ and $m$. 
We assume for simplicity that electrons lie at the lowest energy 
eigenstate characterized by $n=m=0$.
Using the notation $t\equiv t_{m=0}(r,n=0)$, we have 
\begin{eqnarray}
t(r)=\sqrt{\frac{m^*\Omega}{\pi \hbar}}\exp\left(-\frac{m^*\Omega}{2\hbar}r^2\right).
\label{eq:u0} 
\end{eqnarray}
Finally we obtain $\langle L_B \rangle$ in regard to $t$ as 
\begin{eqnarray}
\langle L_B \rangle =\displaystyle\int_0^{\infty}rdr\int_0^{2\pi}
d\theta\ t^*(r) L_B t(r)
=-\frac{\hbar eB}{2m^* \Omega} ,
\label{eq:expectation1} 
\end{eqnarray}
or equivalently, 
\begin{eqnarray}
\frac{\langle L_B \rangle}{\hbar} =\displaystyle-\frac{I_{\rm ext}/I_0}{\sqrt{4+\left(I_{\rm ext}/I_0\right)^2}}\ , \ \ I_0=\frac{m^* \omega_p \ell}{e \mu_0}. 
\label{eq:expectation2}
\end{eqnarray}
From equation (\ref{eq:expectation2}), we see that 
$\langle L_B \rangle \neq 0$ if $I_{\rm ext} \neq 0$. 
This means that a sizeable amount of the quantum phase shift occurs 
by introducing an appropriate magnitude of $I_{\rm ext}$.
\section{Physical consequences}
\subsection{ Persistent current}
An important consequence of the torsion-induced phase shift is the 
occurrence of persistent current in the absence of penetrating 
magnetic flux. 
To derive it, we remind that in a coherent ring (both twisted and untwisted) the $\alpha$th one-particle eigenstate carries the current 
\cite{current}
\begin{eqnarray}
I_{\alpha} =\frac{e \hbar k_{\alpha}}{m^*\ell}. 
\label{eq:dennryuu}
\end{eqnarray} 
The wavenumber $k_{\alpha}$ is determined by considering the periodic 
boundary condition $\psi(q_0 + \ell) = \psi (q_0)$, i.e.,  
\begin{eqnarray}
\exp(-ik\ell) \exp \left( -\frac{i}{\hbar}\tau\langle L_B \rangle \ell 
\right) =1 ,
\label{eq:equivalently}
\end{eqnarray}
which results in 
\begin{eqnarray}
k=\frac{2\pi}{\ell} \alpha - \frac{\tau\langle L_B \rangle}{\hbar} \equiv 
k_{\alpha}. \hspace{2.0mm} (\alpha =0,\pm 1 , \pm 2 \ldots)
\label{eq:wavenumber}
\end{eqnarray}
The total persistent current $I$ carried by the whole electrons 
in a ring is the sum of 
the contributions from all eigenstates. 
From equations (\ref{eq:dennryuu}) and (\ref{eq:wavenumber}), we can 
prove that \cite{taira,62}
\begin{eqnarray}
I=I(p)=\left\{ 
\begin{array}{ccc}
0  &{\rm for}& p=0, \\
\displaystyle 
\frac{ev_F}{\ell}(1-p) &{\rm for}& 0< p <2,
\end{array} 
\right.
\label{eq:matrix}
\end{eqnarray}
where $p=4\tau\langle L_B \rangle \ell/h$ and $I(p)=I(p+2)$. 
The result (\ref{eq:matrix}) indicates that the persistent current 
takes place when $p \neq 0$, i.e., $\langle L_B \rangle \neq0$ that is 
realized by introducing a nonzero $I_{\rm ext}$ as discussed earlier. 

\begin{figure}[t]
\begin{center}
\includegraphics[width=8.5cm]{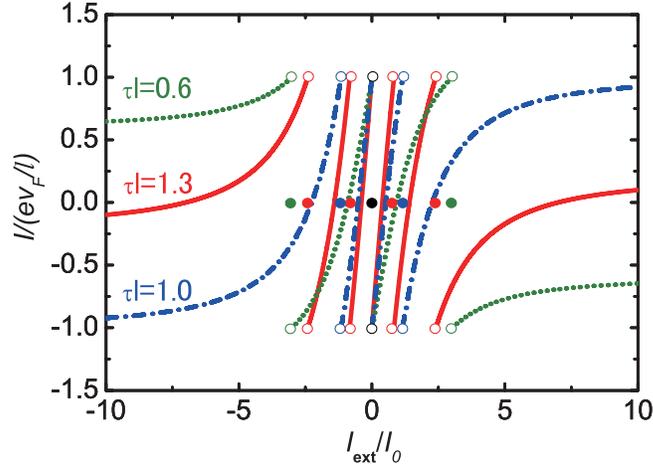}
\caption{\label{fig:fig3} 
Torsion-induced persistent current
$I$ as a function of the external current 
$I_{\rm ext}$ for three different values of torsion $\tau$. 
Open and solid circles represent discrete jumps of $I$ from $-1$ to 
$+1$. 
}
\end{center}
\end{figure}

Figure \ref{fig:fig3} shows the normalized persistent 
current $I/(ev_F/\ell)$ as a function of $I_{\rm ext}$. 
Three different values of the torsion $\tau$ are chosen as indicated 
in the figure. 
With increasing $I_{\rm ext}$, the persistent current $I$ 
jumps from  $-1$ to $0$ and then $+1$ at several discrete points. 
The jump frequently occurs in the region $|I_{\rm ext}/I_0|<4$, while 
it disappears outside the region. 
This non-uniform distribution of discrete jumps is understood by 
equation (\ref{eq:expectation2}); we see from the equation that 
$\langle L_B \rangle$ becomes almost independent of $I_{\rm ext}$ 
in the region $|I_{\rm ext}/I_0| \gg 4$.

\begin{figure}[t]
\begin{center}
\includegraphics[width=16.0cm]{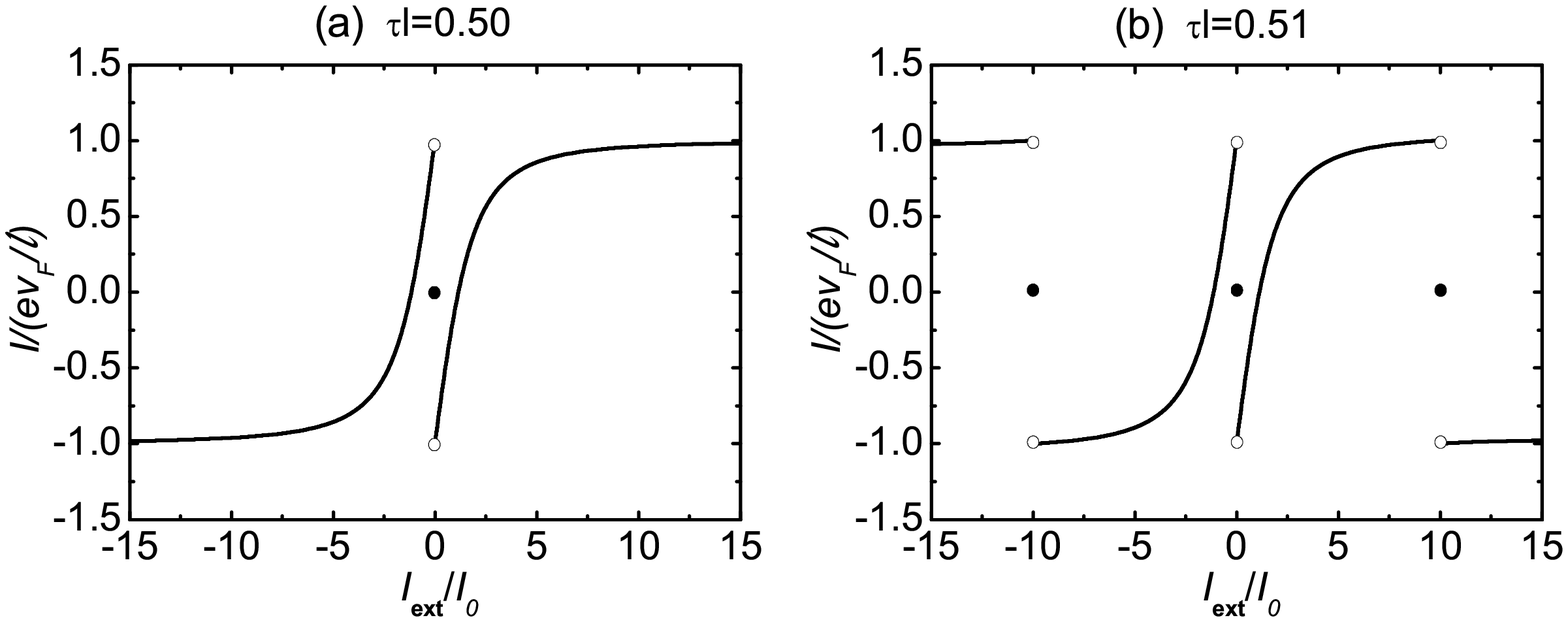}
\vspace{-6.0cm}
\caption{\label{fig:fig4} 
Drastic change in the curve of $I$ across a critical value of the 
torsion $\tau_c$: $\tau<\tau_c$ for (a) and $\tau>\tau_c$ for (b). 
The number of discrete jumps suddenly changes across $\tau_c$.
}
\end{center}
\end{figure}

The position of $I_{\rm ext}$ at which a jump occurs strongly 
depends on $\tau$. 
Figures \ref{fig:fig4}(a) and 
(b) show a drastic change in the curves of 
$I$ across a critical value of $\tau_c$. 
At $\tau < \tau_c$ ($\tau \ell=0.50$), only one jump of $I$ is found 
at $I_{\rm ext}=0$. 
Contrariwise, at $\tau > \tau_c$ ($\tau \ell =0.51$), two additional 
jumps emerge at $I_{\rm ext}/I_0=\pm10$. 
The occurrence of the two jumps at $\tau \ell =0.51$ results from 
the fact that $p=4\tau\ell\langle L_B \rangle/h$ exceeds 
the threshold $p = 2$. 
A further increase in $\tau$ makes the positions of the two jumps 
approach the central 
one at $I_{\rm ext}=0$; it then engenders additional two jumps 
newly at symmetric positions that locate 
outside the existing three jumps. 
It should be emphasized that the discrete jumps in the 
torsion-induced persistent current distribute densely around 
$I_{\rm ext}=0$, whereas those in an ordinary persistent current 
observed in untwisted rings occur periodically in response to an 
increase in the penetrating magnetic flux. 
\subsection{Aharonov-Bohm effect}
Next we discuss an AB-like conductance modulation in the 
electron interferometer based on a twisted quantum ring. 
The interferometer consists of one ring of the 
perimeter $\ell$ and two infinitely long leads 
attached to the opposite sides of the ring. 
Two semicircular arcs of the ring have the same length of $\ell/2$, 
and the incident electron is set to be a plane wave having the 
wavenumber $k$. 
Using the two-terminal Landauer formula \cite{landauer}, we obtain the 
conductance $G$ of the system as 
\begin{eqnarray}
G=\frac{2e^2}{h} \left|(1+\theta_1)\frac{\alpha'\gamma-\beta\gamma'}{\alpha\alpha'-\beta\beta'}+2\theta_1\left( \frac{\alpha \gamma' -\beta' \gamma}{\alpha\alpha'-\beta\beta'}-1\right)\right|^2,
\label{eq:landauer}
\end{eqnarray}
where
\begin{eqnarray}
       \begin{array}{l}
\alpha(k,a) =k\left[2(2k+a)-(k+a)\displaystyle\frac{\theta_2}{\theta_1}+(k-a)\theta_2\right], \\
\vspace{1.0mm}
\beta(k,a) =k\left[2k+a+\displaystyle\frac{a}{\theta_1}+2(k-a)\theta_2\right], \\
\vspace{1.0mm}
\gamma(k,a) =2k[2k+a+(k-a)\theta_2],  \\
\vspace{1.0mm}
\theta_1=\exp(-ik\ell),\ \  \theta_2=\exp(ia\ell),\ \  a=\displaystyle\frac{\tau \langle L_B \rangle}{\hbar},
       \end{array}
\label{eq:abc} 
\end{eqnarray}
with a notation $\xi'(k,a)=\xi(k,-a)$ for $\xi=\alpha,\beta,\gamma$. 
Note that $G$ is dependent on the dimensionless parameters 
$k \ell$ and $a \ell$, or equivalently, $k$, $\tau$ and 
$I_{\rm ext}$. 
Among choices, we consider the behavior of $G$ in the following 
two cases: the dependence of $G$ on $\tau$ and $I_{\rm ext}$ 
($k$ is fixed), and that of $k$ and $I_{\rm ext}$ ($\tau$ is fixed).

Figures \ref{fig:fig5}(a) and (b) show three-dimensional 
plots of the dimensionless conductance $\tilde{G} \equiv G/(2e^2/h)$. 
The two plots exhibit the modulation of $\tilde{G}$ in response to 
changes in the parameters: $k$, $\tau$, and $I_{\rm ext}$. 
We emphasize that the modulation of $G$ requires 
no magnetic flux threading the twisted ring. 
The degree of modulation is pronounced close to the line of 
$I_{\rm ext}=0$ around which the surface of $\tilde{G}$ is corrugated 
steeply. 
This pronounced modulation close to $I_{\rm ext}=0$ is similar to the 
dense population of jumps observed in the persistent 
current around $I_{\rm ext}=0$. 
In fact, both features are attributed to the nonlinear behavior of 
$\langle L_B \rangle$ as a function of $I_{\rm ext}$ (See equation 
(\ref{eq:expectation2})).

\section{Discussion}
We remarks another possible way to obtain a nonzero 
value of $\langle L_B \rangle$. 
Instead of the introduction of $I_{\rm ext}$, we may directly 
apply an external magnetic field in 
a direction {\it tangential} to a twisted structure. 
Such the tangential field plays a similar role to $I_{\rm ext}$ 
depicted in figure \ref{fig:fig2}, and therefore, it causes 
torsion-induced current flow and the conductance modulation in the 
twisted ring. 

\begin{figure}
\centering
\subfigure{
    \resizebox{9.5cm}{!}{\includegraphics{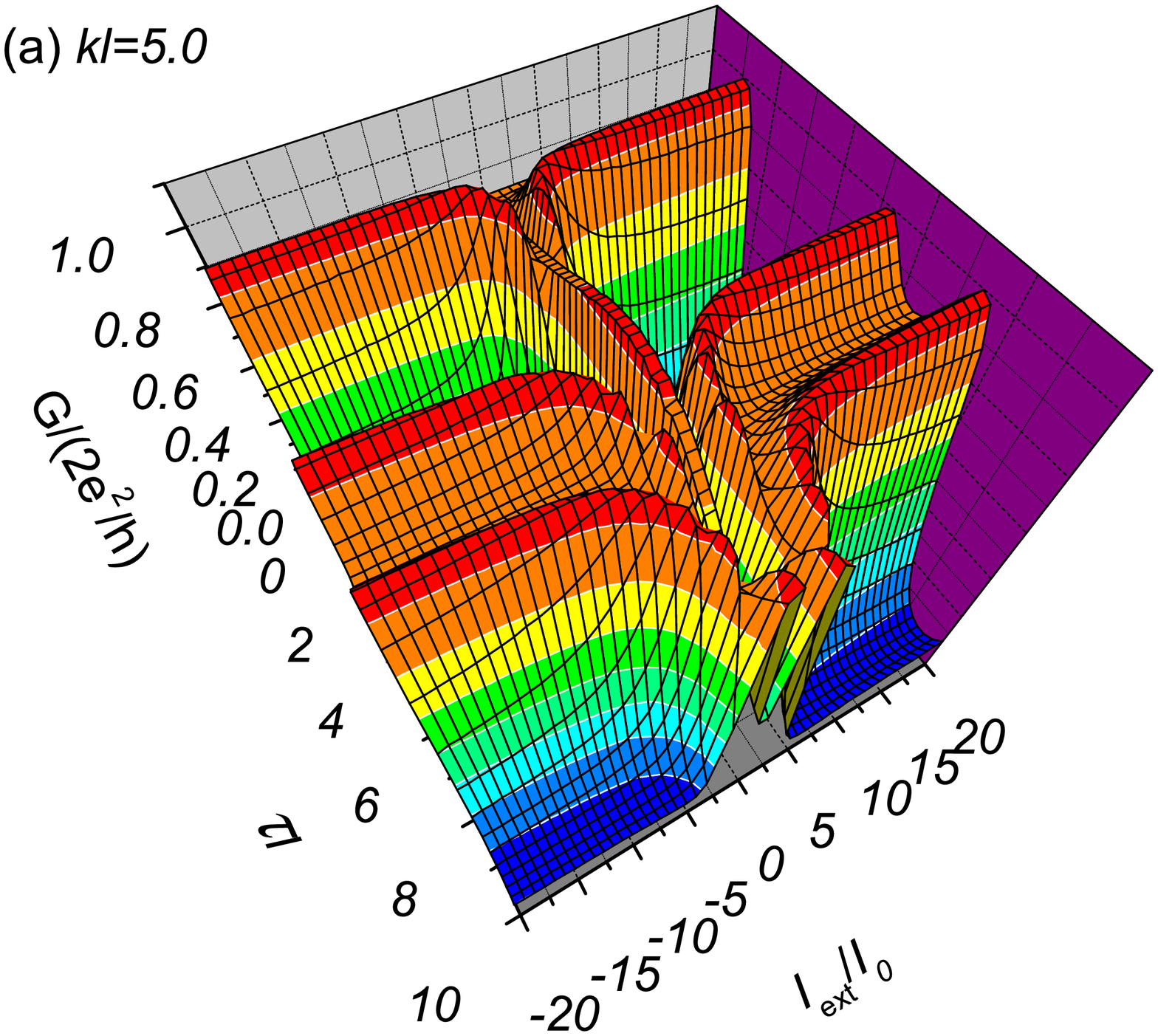}}
\hspace{-2.0cm}
    \resizebox{9.5cm}{!}{\includegraphics{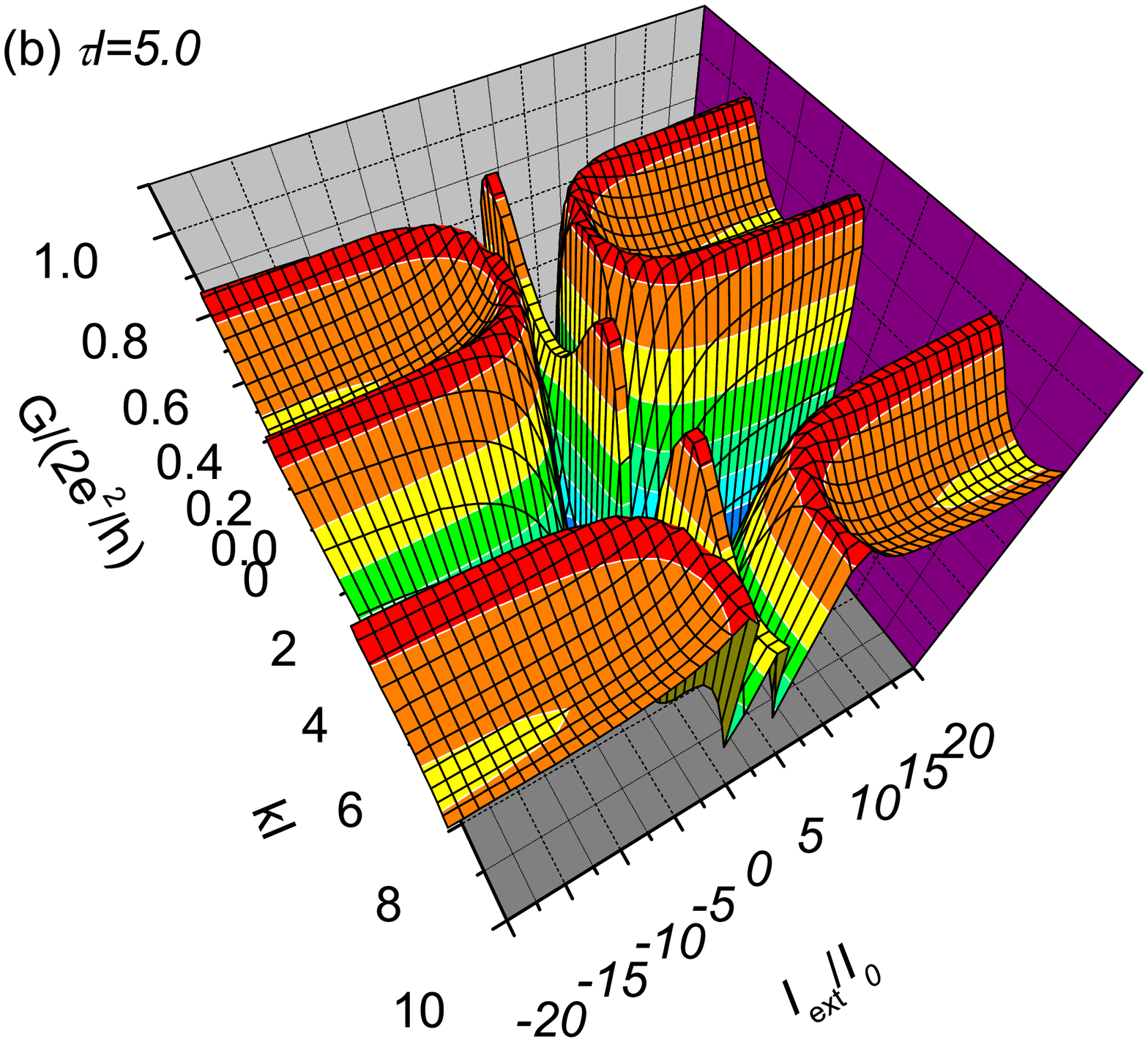}}
}
\caption{
Three-dimensional plots of the dimensionless conductance 
$\tilde{G}/(2e^2/h)$ on the two-parameter spaces: $I_{\rm ext}/I_0$ 
and $\tau \ell$ in (a); $I_{\rm ext}/I_0$ and $k\ell$ in (b).
}
\label{fig:fig5}
\end{figure}

It is also noteworthy that our results of the persistent current (figure \ref{fig:fig4}) and conductance oscillation (figure \ref{fig:fig5}) should be dependent on the ring's parameters: the ring's perimeter ($=\ell$) and its cross-sectional area (characterized by $\omega_p$) are cases in point. These dependences are partly described by equation (\ref{eq:expectation2}); it shows that a sizable amount of the $I_{\rm ext}$-driven phase shift is obtained only when $|I_{\rm ext}/I_0| < 4$, in which $I_0 \propto \ell \omega_p$. Hence, a smaller $\ell$ or $\omega_p$ requires a larger $I_{\rm ext}$ for obtaining the same torsion-induced phase shift. In addition, if $\omega_p$ is too small, the cross-sectional area becomes so large and thus the system can be no longer considered as 1D, since quantum excitations in the transverse directions are allowed. The effect of thickness would be trivial if the ring were untwisted; the amplitudes of the conductance \cite{PRLAB} and persistent current \cite{PRLPC} will increase, since the number of conducting channels increases and all the channels assume the same amount of phase shift. Contrariwise, in a twisted ring with finite thickness, each conducting channel assumes different values of the phase shift \cite{torsion3}. Therefore, further study is needed to solve the thickness effect on the persistent current and conductance modulation in twisted rings.

%
\section{Summary}
We have demonstrated a novel type of quantum phase shift induced by 
internal torsion of quasi-1D twisted rings. 
The degree of the phase shift is proportional to torsion $\tau$ and the 
ring perimeter $\ell$, and shows a nonlinear response to the 
external current flow $I_{\rm ext}$ threads inside the ring. 
This phase shift drives the persistent current $I$ and the AB-like 
oscillation in the conductance $G$. 
Detailed analyses of the non-periodic responses of $I$ and $G$ to $I_{\rm ext}$ proved the distinct properties of the two phenomena from the ordinary 
counterparts appearing in untwisted rings. 

\ack
We wish to express our thanks to K Yakubo for illuminating discussions and 
M Arroyo for his help and hospitality in using the
facility of UPC.
HT acknowledges the financial supports 
from the JSPS Research Fellowships for Young Scientists and the
Excellent Young Researcher Overseas Visit Program.
HS thanks the supports by the Kazima Foundation and a Grant-in-Aid 
for Scientific Research from the MEXT, Japan. 
A part of numerical calculations were performed using facilities of the Institute for Solid State Physics, University of Tokyo.


\end{document}